\documentclass[aps,prb,showpacs,amsmath,amssymb,graphicx,twocolumn]{revtex4}
\usepackage{graphicx}
\usepackage{bm}
\begin{document}
\title{Dissipation and coherent effects in narrow superconducting channels}
\author{M. Amin Kayali$^{1,4}$, Vladimir G. Kogan$^{3}$ and
Valery L. Pokrovsky$^{1,2}$}
\affiliation{$^1$Department of Physics,
Texas A \& M University College Station, Texas 77843-4242, USA.\\
$^2$Landau Institute for Theoretical Physics, Chernogolovka, Moscow,
142432, Russia.\\ $^3$ Ames Laboratory-DOE and Department of
Physics, Iowa State University, Ames, Iowa 50011\\$^4$ Human
Neuroimaging Lab., Baylor College of Medicine, One Baylor Plaza,
Houston, TX 77030}
\date{\today}
\begin{abstract}
We apply the time dependent Ginzburg-Landau equations (TDGL) to study small ac
currents of  frequency $\omega$  in  superconducting  channels narrow on
the scale of London penetration depth. We show that TDGL have $t$-dependent
and spatially uniform solutions that describe the order parameter
with an oscillating part of the double frequency coexisting with an ac
electric field. We evaluate the Ohmic losses (related  neither to the flux
flow nor to the phase slips)  and show that  the resistivity reduction
on cooling through the critical temperature $T_c$ should behave as
$(T_c-T)^2/\omega^2$. If the channel is cut out of an anisotropic
material in a direction other than the principal axes, the transverse 
phase difference and the Josephson voltage between the channel sides 
are generated.
\end{abstract}
\noindent 
\pacs{74.20.-z,74.20.De,74.50.+r}
\maketitle
\section{Introduction}

It is a common knowledge that superconductors dissipate in the presence of the
flux flow or, for large driving current densities, due to phase slips. It is
also known that even a small ac current in zero applied field causes
dissipation when none of the above sources of dissipation are present.  E.g.,
the resistive transition to the superconducting state recorded with small ac
currents has always a finite width which for small enough currents and in zero
field not always can  be relegated to the flux flow, sample inhomogeneities,
or  thermal fluctuations. A qualitative explanation of this dissipation 
employs the two-fluid model with the normal and superfluid densities, $n_n$ 
and $n_s$, constant in space and time.\cite{Tinkham}

Ohmic losses in superconductors are absent for small dc currents.
As  was originally argued by Landau for superfluids, the flow of
quasiparticles is stopped by the lattice (phonons) or by impurities and does
not contribute to the current, whereas the creation of new excitations is
prohibited by the gap in the quasiparticle spectrum.
The situation is different for ac currents. During the ac period 
$2\pi/\omega$, the normal part of the Fermi-liquid does not  stop completely  
and, therefore, causes Ohmic losses.
When $\omega $ significantly exceeds the phase relaxation rate $\tau
_{J}^{-1}$, but still is small relative to the normal carriers relaxation rate
$\tau _{n}^{-1}$, the Ohmic losses should approach their normal limit
$\sim J^{2}/\sigma $.

These results were obtained within microscopic theory, see e.g.,
Ref.\,\onlinecite{Bardeen}.
In this article we show that for low frequencies  the TDGL offers a general
and simple method to approach the dissipation problem near the transition
point without specific assumptions on the dissipation mechanism. We
show that if superconducting wires (channels) are thin compared to the London
depth $\lambda$ and the ac current can be taken as uniform, the order parameter
acquires a part oscillating in time with the frequency $2\omega$ where $\omega$
is the current frequency. The order parameter modulus stays constant in
space since no vortices or phase slips are assumed to exist in zero applied
field and for sufficiently small currents. One can say that there is a periodic
exchange between the superfluid condensate and the normal excitations,
accompanied by an ac electric field $E$. In general, the phase shift between
the field and the current depends on relative values of
$\omega\tau_\Delta$ and $\omega\tau_J$ with $\tau_\Delta$ and $\tau_J$ being
the  relaxation times for the order parameter  and for the phase (i.e., 
for the current). As a consequence, the dissipation depends on these
parameters, too.

In anisotropic superconducting channels, the ac currents flowing in any
but the principal crystal directions cause the electric field to have a
component perpendicular to the current, i.e., across the channel. This is due
to the  anisotropy of the superconductor in use and due to anisotropy of the
normal conductivity. We show that for $\omega\tau_J\ll 1$ the transverse
field is caused by the inherent for anisotropic superconductors
transverse phase difference.  This offers a relatively simple probe of
existence of this phase difference which has been recently predicted.\cite{KP}

\section{Isotropic case}

To set notations, we start with the first GL equation:
\begin{equation}
-\xi^2  {\bm \Pi}^2 \psi=\psi
\left(1-|\psi|^2/\psi^2_0\right)\,.
\label{GL}
\end{equation}
Here, $\xi$ is the coherence length, ${\bm \Pi}=\nabla+2\pi i{\bm A}/\phi_0$ 
with ${\bm A}$ and $\phi_0$ being the vector potential and  the flux quantum.
For the order parameter written as $\psi=f\,e^{i\chi} $, we have
${\bm \Pi}\psi =e^{i\chi}(\nabla f+i{\bm P}f)$ where ${\bm P}$ is 
proportional to the gauge invariant vector potential
\begin{equation}
{\bm Q}={\phi_0\bm \nabla}\chi/2\pi+{\bm A}=\phi_0{\bm P}/2\pi \,.
\end{equation}

Equation (\ref{GL}) contains $e^{i\chi}$ on both sides. After cancelling 
this factor and separating real and imaginary parts, one obtains for the real 
part:
\begin{eqnarray}
-\xi^2   ( \nabla^2f -f\,P^2 )=f(1-f^2/f_0^2)\,. \label{GL1}
\label{GL1.1}
\end{eqnarray}
The imaginary part coincides with div${\bm j}={\rm div}f^2 {\bm P=0}$. The 
gauge invariant form of TDGL involves the scalar potential
$\varphi$:\cite{Schmid,Kopnin,Gorkov-Kopnin}
\begin{equation}
\tau_\Delta \left(\frac{\partial }{\partial t} -i\frac{2\pi
c}{\phi_0}\,\varphi\right)\psi = \psi
\left(1-\frac{f^2}{f^2_0}\right) +  \xi^2  {\bm \Pi}^2\psi\,,
\label{GLan}
\end{equation}
where $\tau_\Delta$ is the order parameter  relaxation time. Separating 
real and imaginary parts we have:
\begin{eqnarray}
\tau_\Delta \frac{\partial f}{\partial t} &=&f\left(1-\frac{f^2}{f_0^2}\right)
+\xi^2 \left (\nabla^2f-fP^2 \right ), \label{TDGL1}\\
-\tau_\Delta c  f\Phi
&=&  \nabla f\cdot{\bm Q} + \nabla\cdot(f{\bm Q} )   \,, \label{TDGL2}
\end{eqnarray}
where $\Phi=\varphi-(\phi_0/2\pi c)\partial_t\chi$.

\section{Spatially uniform solutions of TDGL}

We are interested in coordinate independent solutions $f(t)$ and $Q_i(t)$.
The system (\ref{TDGL1}),\,(\ref{TDGL2}) then takes the form:
\begin{eqnarray}
{\tau_\Delta\over 2} \frac{\partial u}{\partial t}
&=&u\left(1-u  \right) -\xi^2  P^2\, u \,,\quad
u=\frac{f^2}{f_0^2}\,,\label{eq.for_f} \label{u}\\
\Phi&=&0\,.\label{Phi}
\end{eqnarray}
This  should be complemented with equations for the current. A uniform 
current ${\bm J}$  consists, in general, of  normal and superconducting parts:
\begin{equation}
J= \sigma E -\frac{2 e^2}{Mc}\,f^2 Q \,,
\end{equation}
where $ E$ is the electric field directed along the channel and $\sigma$ 
is the conductivity  for the quasiparticles flow. We aim to describe the 
system response to ac currents; then $\sigma$ is in general 
$\omega$-dependent. If however the frequencies   are bound by inequality 
$\omega\tau_n\ll 1$ with $\tau_n$ being the scattering time for the normal 
excitations, one can consider $\sigma$ as a real $\omega$-independent quantity.

The electric field is expressed in terms of gauge invariant
potentials:
\begin{equation}
{\bm E}= -\nabla\Phi- \frac{\partial {\bm Q}}{c\partial t} = -
\frac{\partial {\bm Q}}{c\partial t}\,,
\label{E}
\end{equation}
so that the total current is
\begin{equation}
J =  -\frac{1}{c}\left(\sigma \frac{\partial
Q }{\partial t}+\frac{2 e^2}{M}\,f^2Q \right)\,.\label{jx}
\end{equation}

At a given current, Eqs.\,(\ref{u}) and (\ref{jx}) form a complete
system for two functions $u(t)$ and $Q(t)$. It is convenient to 
introduce dimensionless vector potential
\begin{equation}
q=Q \,\frac{2\pi\xi}{\phi_0}
\end{equation}
and to measure the current density in units of the depairing value:
\cite{remark} $j=J/J_{GL}$ with
\begin{equation}
J_{GL}=\frac{e^2\phi_0f_0^2}{ \pi\xi
cM}=\frac{c\phi_0}{4\pi^2\xi\lambda^2}\,,\quad
\lambda^2=\frac{Mc^2}{4\pi e^2f_0^2}\,. \label{jdp}
\end{equation}
Also, we use the so-called ``current relaxation time"
\begin{equation}
\tau_J=\frac{M\sigma}{2  e^2f_0^2}\sim\frac{n_n}{n_s}\,\tau_n\,.
\label{tau_j}
\end{equation}
In our view, a better term for this quantity is the ``phase relaxation time"
which we use in what follows, but we keep the standard notation $\tau_J$.  
When $T\to T_c$, $\tau_J\propto 1/(T_c-T)$  and so does $\tau_\Delta$.
\cite{Kopnin}

Then, the system of equations to solve takes the form:
\begin{eqnarray}
&& \tau_\Delta \, {\dot u} /2 = u -u^2   -   q^2 u \,,   \label{is1}\\
&&  -\tau_J\, {\dot q}  -u q = j   \,,\label{is2}
\end{eqnarray}
where dots stand for $d/dt$.

We are interested in calculating the system response to ac currents
$J=J_0\cos\omega t$ with amplitudes $J_0\ll J_{GL}$, i.e., for $j\ll
1$. In this situation the order parameter $u$ is close to unity and
$q\ll 1$. The system to solve can be simplified ($u\approx 1-v$,
$v\ll 1$):
\begin{eqnarray}
&& \tau_\Delta\,\dot{v}/2 + v = q^2   \,,   \label{is1l}\\
&& \tau_J\,\dot{q} + q =-j_0\cos\omega t \,.\label{is2l}
\end{eqnarray}
The second equation here is {\it linear}; moreover, it is decoupled from 
the equation for $v$ and is easily solved. The solution consists of a 
transient part depending on initial conditions (the general solution of 
the homogeneous equation) and the long time asymptotics of
our interest (the particular solution of the full equation). The latter can be
readily found by looking for $q$ of the form $A \sin\omega t+B \cos\omega t$:
\begin{equation}
q(t\to\infty)=-\frac{j_0(\omega \tau_J\sin\omega t+\cos\omega
t)}{1+\omega^2\tau_J^2}\,.
\label{q_is}
\end{equation}
This can also be written in a more familiar form:
\begin{equation}
q_\infty=-\frac{j_0  }{\sqrt{1+\omega^2\tau_J^2}}\,\sin(\omega
t+\alpha)\,,\quad
\tan\alpha=\frac{1}{\omega \tau_J}.
\label{q_is}
\end{equation}
In the following we are interested only in the stationary long time asymptotics
and  omit the subscript $\infty$. Substituting the obtained $q$ in 
Eq.\,(\ref{is1l}) we can find  the long time asymptotics for $v$. To this 
end, we look for $v=v_0+v_1\cos 2\omega t+v_2\sin 2\omega t$  and obtain:
\begin{eqnarray}
&&v_0=\frac{j_0^2}{2(1+\omega^2\tau_J^2)}\,,\\
&&v_1=\frac{j_0^2(1-\omega^2\tau_J^2-2\omega^2\tau_J
\tau_\Delta)}{2(1+\omega^2\tau_J^2)^2(1+\omega^2\tau_\Delta^2)}\,, \\
&&v_2= \frac{j_0^2\omega[2\tau_J + \tau_\Delta(1-\omega^2\tau_J^2)]}
{2(1+\omega^2\tau_J^2)^2(1+\omega^2\tau_\Delta^2)}\,.
\end{eqnarray}
This yields:
\begin{eqnarray}
&&u =1-\frac{j_0^2}{2(1+\omega^2\tau_J^2)} -\frac{j_0^2}{2}
   \sin(2\omega t+ \beta)\,,\label{u_ass}\\
&&\tan\beta=\frac{v_1}{v_2}\,.
\end{eqnarray}

Thus, in the stationary state reached when
$t\gg{\rm max}(\tau_\Delta,\tau_J)$, the order parameter has a part oscillating
with frequency $2\omega$ near the average value given in the first two terms of
Eq.\,(\ref{u_ass}). Clearly, the frequency doubling is due to the
order parameter independence on the current direction. The zero frequency 
limit of Eq.\,(\ref{u_ass}) coincides with the known GL result for the order 
parameter suppression by a dc current: $u=1-j_0^2$.

Oscillations of the order parameter {\it per se} are difficult to measure. 
This is not the case for the electric field $E$ and the dissipation density 
$W=JE$.
The field $E$ of Eq.\,(\ref{E}) in the stationary long time state is
\begin{equation}
E=-\frac{\phi_0{\dot q}}{2\pi\xi c}=\frac{\phi_0}{2\pi\xi
c}\,\frac{j_0\omega}{\sqrt{1+\omega^2\tau_J^2}}\cos(\omega t+\alpha)\,.
\end{equation}
The dissipation averaged over the oscillations period follows:
\begin{equation}
\overline{W} =
\frac{\pi J_0^2 \lambda^2\omega^2\tau_J}{ c^2 (1+\omega^2\tau_J^2)}\,.
\label{W}
\end{equation}
It is worth noting that for small currents both the electric field
and  dissipation are not affected by the order parameter relaxation time 
$\tau_\Delta$. For $\omega\tau_J\ll 1$, the field $E\propto \omega$ and 
$\overline{W}\propto\omega^2$; they are $\omega$-independent for
$\omega\tau_J\gg 1$.

Since $\tau_J$ diverges when $T\to T_c$, we obtain in this limit
the dissipation in the normal state $\overline{W} =J_0^2/2\sigma$, as
expected. Expanding $\overline{W}$ of Eq.\,(\ref{W}) in the small parameter
$1/\omega^2\tau_J^2$ and keeping the first correction  we obtain
\begin{equation}
\overline{W} \approx
\frac{ J_0^2}{2\sigma}\left(1-\frac{4e^4f_0^4}{M^2\sigma^2\omega^2}\right)  \,.
\label{W_av}
\end{equation}

While looking at the $T$-dependence of the dissipation near $T_c$, it should be
noted that the quasiparticle conductivity $\sigma=n_ne^2\tau_n/M$ decreases 
linearly in $(T_c-T)$ due to a decrease of the normal density $n_n$. 
This causes an initial increase of $\overline{W}$, which can be considered 
as manifestation of well-studied coherence effects in electromagnetic 
absorption. However, for frequencies of our interest $\omega\tau_n\ll 1$, 
the ``bump" (the maximum) in the dissipation $\Delta \overline{W} 
\sim (J_0^2/\sigma)\,\omega^2\tau_n^2$ is situated at $ T\approx 
T_c(1- \omega^2\tau_n^2)$, i.e., very close to $T_c$. Out of this narrow 
temperature domain  the dissipation  reduction on cooling through $T_c$ 
should behave as $(T_c-T)^2/\omega^2$.

\section{Anisotropic channel}

In isotropic superconductors in the presence of persistent currents, the 
gauge invariant gradient $\nabla\Theta=\nabla\chi + 2\pi{\bm A}/\phi_0$  
is directed along the current. Recently Kogan and Pokrovsky showed that in 
anisotropic superconductors the transverse phase difference may appear if the
driving current does not point in any of the principal crystal
directions.\cite{KP} In particular, this situation is realized in
current carrying channels cut out of anisotropic crystals with a
long side in any but a principal direction and which are narrow on
the scale $\lambda$. One of the possible ways to observe the transverse phase
is to measure the voltage  $V$ generated by time-dependent phase difference
according to the Josephson formula $V=(\hbar/2e)\partial \Delta
\Theta/\partial t$. This can be achieved by driving an ac current through
the said channel, a simpler possibility to observe the transverse phase than
that suggested in Ref. \onlinecite{KP}.

In the static case, the supercurrent density is given by
$J_i=2e\hbar M_{ik}^{-1} |\Delta|^2 P_k$, where $M_{ij}$ is the 
superconducting mass tensor; the summation is implied over repeated indices. 
It is convenient to introduce the normalized inverse mass tensor 
$\mu_{ik}=M_{ik}^{-1}M$ with $M=(M_a M_b M_c)^{1/3} $; then the eigenvalues 
are  related by $\mu_a\mu_b\mu_c =1$. In the uniaxial case, 
$\mu_a^2 \mu_c=1$, the inverse masses can be expressed in terms of a single 
anisotropy parameter $\gamma^2= \mu_a/\mu_c$: $\mu_a= \gamma^{2/3},\,\,
\mu_c=\gamma^{-4/3}$.

\begin{figure}[ht]
   \vspace{0.1in}
   \centering
   \includegraphics[angle=0,width=3.in]{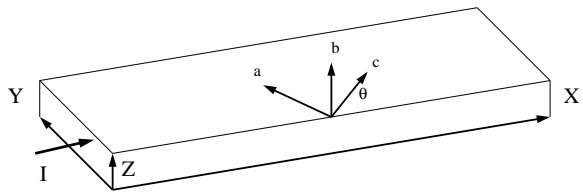}
   \caption{The superconducting channel with a long dimension along $\bm
x$, the direction of an ac current. The
crystal axis ${\bm{b}}$ is directed along $\bm z$. The other crystal  axes
   are in the $xy$ plane with a misalignment angle $\theta$
between ${\bm{c}}$ and $\bm x$.}
   \label{fig1}
\end{figure}

In the coordinates of Fig.\ref{fig1}, the components $\mu_{ik}$ are:
\begin{eqnarray}
\mu_{xx}&=&\gamma^{-4/3}(\gamma^2\sin^2 \theta + \cos^2 \theta)\,,
\nonumber \\
\mu_{yy}&=&\gamma^{-4/3}(\gamma^2\cos^2 \theta +  \sin^2 \theta)\,,
\label{eq2}\\
\mu_{xy}&=&\gamma^{-4/3}(1-\gamma^2)\sin \theta \cos \theta\,,
\nonumber
\end{eqnarray}
whereas $\mu_{zz}=\mu_b=\gamma^{2/3}$ and $\mu_{zx}=\mu_{zy}=0$.

To describe $t$-dependent situations, we again employ the TDGL model, the
generalization of which to the anisotropic situation is straightforward: one
has to replace the operator $\xi^2 {\bm \Pi}^2$ in Eq.\,(\ref{GL}) with
$\xi^2\mu_{ik} \Pi_i\Pi_k$ (see, e.g., Ref.\,\onlinecite{kogan2002}). Then, we
employ the same procedure as in the isotropic case to make the model
dimensionless. The scalar quantities $\xi$ and $\lambda$ now have meaning 
of averages $(\xi_a\xi_b\xi_c)^{1/3}$ and 
$(\lambda_a\lambda_b\lambda_c)^{1/3}$, respectively.

As a result, the system (\ref{is1}),\,(\ref{is2}) is replaced with
\begin{eqnarray}
&& \tau_\Delta {\dot u }/2 = u-u^2 -  \mu_{ik} q_iq_k\, u  \,,
\label{anis1 }\\
&& \tau_Js_{ik} {\dot q_k } +  u\mu_{ik}q_k=-j_{0}\delta_{ix}\cos\omega t\,.
\label{anis2 }
\end{eqnarray}

The tensor $s_{ik}=\sigma_{ik}/\sigma$ is the normalized conductivity with
$\sigma=(\sigma_a \sigma_b\sigma_c)^{1/3}$.
As with the mass tensor, we can introduce for the uniaxial case the
conductivity anisotropy parameter $\gamma_{\sigma}^2=\sigma_a/\sigma_c$
so that $s_a=s_b=\gamma_{\sigma}^{2/3}$ and
$s_c=\gamma_{\sigma}^{-4/3}$.
With these definitions, the components of $s_{ik}$ are given by
formulas (\ref{eq2}) in which $\gamma$ is replaced with
$\gamma_{\sigma}$.

For small currents, we have for $v= 1-u $:
\begin{eqnarray}
&& -\tau_\Delta {\dot v}/2 = v   -  \mu_{ik} q_iq_k   \,,
\label{anis1l}\\
&& \tau_Js_{ik} {\dot q_k} +  \mu_{ik}q_k
=-j_{0}\delta_{ix}\cos\omega t
\,,\label{anis2l}
\end{eqnarray}
As in the isotropic case, the equation for $q_i$ is decoupled from
that for $v$. One can look for a particular solution of Eq.\,(\ref{anis2l}) 
in the form $q_i=A_i\sin\omega t+B_i\cos\omega t$ to obtain a linear 
system of equations for $A_i,B_i$.

Perhaps, the easiest is to deal with Eq.\,(\ref{anis2l}) in the
crystal frame $(a,b,c)$ where all material tensors are diagonal. In this
frame, the equation to solve reads:
\begin{eqnarray}
\tau_Js_\alpha {\dot q_\alpha} +  \mu_\alpha q_\alpha
=-j_{0\alpha} \cos\omega t\,,\quad \alpha=a,c\,;\nonumber\\
j_{0a}=-j_0\sin\theta\,,\qquad j_{0c}=j_0\cos\theta
\,.\label{alpha}
\end{eqnarray}
The long time asymptotics is easily obtained:
\begin{equation}
q_\alpha
=-\frac{j_{0\alpha}\sin(\omega t+\beta_\alpha)}{\sqrt{
\omega^2\tau_J^2s_\alpha^2+\mu_\alpha^2}}\,\,,\quad
\tan\beta_\alpha=\frac{\mu_\alpha}{\omega \tau_Js_\alpha}.
\label{q_is}
\end{equation}
The electric field components in the channel frame $(x,y)$ and the dissipation
read:
\begin{eqnarray}
E_x&=&\frac{\phi_0j_0\omega}{2\pi\xi c}\left[\frac{\sin^2\theta\cos(\omega
t+\beta_a)}{\sqrt{\omega^2\tau_J^2s_a^2+\mu_a^2}} +
\frac{\cos^2\theta\cos(\omega
t+\beta_c)}{\sqrt{\omega^2\tau_J^2s_c^2+\mu_c^2}}\right],\nonumber\\
E_y&=&\frac{\phi_0j_0\omega\sin2\theta}{4\pi\xi
c}\left[\frac{ \cos(\omega
t+\beta_c)}{\sqrt{\omega^2\tau_J^2s_c^2+\mu_c^2}}-\frac{ \cos(\omega
t+\beta_a)}{\sqrt{\omega^2\tau_J^2s_a^2+\mu_a^2}}
\right],\nonumber\\
\overline{W}&=&
\frac{\pi J_0^2 \lambda^2\omega^2\tau_J}{
c^2}\left[\frac{s_a\sin^2\theta}{\omega^2\tau_J^2s_a^2+\mu_a^2}+
\frac{s_c\cos^2\theta}{\omega^2\tau_J^2s_c^2+\mu_c^2}\right].
\label{general}
\end{eqnarray}
Clearly, these expressions have the correct isotropic limit.
It is instructive to consider a few limiting situations.

1. For a dc current ($\omega=0$), Eq.\,(\ref{anis1l}) gives $v=\mu_{ik}
q_iq_k$ whereas Eq.\,(\ref{anis2l}) yields $q_i=-j_0\mu_{xi}^{-1}$. We then
have
\begin{equation}
u=1-j_0^2\mu_{xx}^{-1}\,,
\end{equation}
so that the order parameter suppression by a dc current depends on the current
direction. We will not write down a cumbersome expression for  $u$ in
the general case, but the physics here is the same as for
the isotropic case: the order
parameter has a small part oscillating with frequency $2\omega$.\\

2. $\omega \tau_J\ll 1$. This situation corresponds to temperatures not too
close to the critical temperature because $\tau_J\to\infty$ when $T\to T_c$.
We have:
\begin{equation}
  \overline{W}=  \frac{ J_0^2 }{2\sigma}\,
\omega^2\tau_J^2\left(\frac{s_a}{\mu_a^2}\sin^2\theta
+\frac{s_c}{\mu_c^2}\cos^2\theta\right) \,,
\end{equation}

In the linear approximation in the small $\omega
\tau_J$, the time averaged   dissipation is absent; the energy during each
period is pumped from the condensate to the normal excitations and back in
equal amounts. The electric fields are:
\begin{equation}
E_x= \frac{\phi_0\omega j_0}{2\pi\xi c}\,\mu_{xx}^{-1}\sin\omega t,\quad
E_y=E_x\,\frac{\mu_{xy}^{-1}}{\mu_{xx}^{-1}}.
\label{eq39}
  \end{equation}
The conductivity tensor $s_{ik}$ does not enter these expressions. One may say
that these electric fields are due to the $t$-dependence of the phase
differences (the Josephson relation mentioned above). In
particular, the very fact that the transverse field $E_y\ne 0$ is a proof of 
existence of the transverse phase. Hence, measuring the transverse and
longitudinal voltages on a channel similar to the shown in Fig.\,\ref{fig1},
one can, in principle, verify Eq.\,(\ref{eq39}) and therefore observe the
transverse phase difference.
\\

3. $\omega \tau_J\gg 1$, the situation taking place in particular when $T\to
T_c$. The dissipation of Eq.\,(\ref{general}) reduces to  the
form similar to that of the isotropic case:
\begin{equation}
\overline{W} \approx   \frac{
J_0^2}{2}\,\sigma_{xx}^{-1}\left(1-\eta(\theta)\frac{4e^4f_0^4}
{M^2\sigma^2\omega^2}\right)
\end{equation}
with
\begin{equation}
\eta=  (\gamma_\sigma \gamma^2)^{-4/3}\,
\frac{ \gamma^4\sin^2\theta+ \gamma_\sigma^6\cos^2\theta}{\sin^2\theta+
\gamma_\sigma^2\cos^2\theta } \,.
\end{equation}
At $T_c$, $\overline{W} =J_0^2\sigma_{xx}^{-1} /2 $ is the normal state
dissipation. Thus, on cooling through $T_c$, the resistivity drop should
behave as $(T_c-T)^2/\omega^2$ with an angular dependent coefficient. It
should be noted that $\gamma_\sigma$ may exceed substantially the
superconducting anisotropy $\gamma$ causing  a strong angular dependence of
$\eta$.

We have also performed the linear stability analysis of our solutions of TDGL
to show that the homogeneous solution is stable unless the
current reaches the magnitude of the order of $J_{GL}$. One can argue that
vortices might be generated near the boundaries at smaller currents thus
destroying the uniform time-dependent states. Without going to a detail
discussion of this restriction, we note that our solutions  for small
currents are certainly stable. \\

This work was supported by the NSF grant DMR 0321572 and DOE grant
DE-FG03-96ER45598, and Telecommunication and Information Task Force
at Texas A \& M University. The work of V.K. is supported by the
Office of Basic Energy Sciences of the U. S. Department of Energy;
Ames Laboratory is operated for DOE by Iowa State University under
Contract No. W-7405-Eng-82.


\begin{thebibliography}{99}
\bibitem{remark}For convenience, we   define  the depairing current $J_{GL}$ as
to have the order parameter suppression by  small currents to be
$f^2/f_0^2=1-J^2/J_{GL}^2$. This definition differs by
a constant factor from a standard definition of $J_{GL}$ as a maximum value for
which the GL equations still have stable solutions in the presence of large
currents.\cite{Tinkham}
\end{thebibliography}
\end{document}